# Predicting non-neutral missense mutations and their biochemical consequences using genome-scale homology modeling of human protein complexes

Andrew J. Bordner* and Barry Zorman
Mayo Clinic, 13400 East Shea Boulevard, Scottsdale, AZ  85259  USA

## Abstract
Computational methods are needed to differentiate the small fraction of missense mutations that contribute to disease by disrupting protein function from neutral variants. We describe several complementary methods using large-scale homology modeling of human protein complexes to detect non-neutral mutations. Importantly, unlike sequence conservation-based methods, this structure-based approach provides experimentally testable biochemical mechanisms for mutations in disease. Specifically, we infer metal ion, small molecule, protein-protein, and nucleic acid binding sites by homology and find that disease-associated missense mutations are more prevalent in each class of binding site than are neutral mutations. Importantly, our approach identifies considerably more binding sites than those annotated in the RefSeq database. Furthermore, an analysis of metal ion and protein-protein binding sites predicted by machine learning shows a similar preponderance of disease-associated mutations in these sites. We also derive a statistical score for predicting how mutations affect metal ion binding and find many dbSNP mutations that likely disrupt ion binding but were not previously considered deleterious. We also cluster mutations in the protein structure to discover putative functional regions. Finally, we develop a machine learning predictor for detecting disease-associated missense mutations and show that it outperforms two other prediction methods on an independent test set.

## Author Summary
Mutations affecting proteins are responsible for causing most inherited human diseases as well as contributing to the initiation and progression of cancer. New DNA sequencing technologies can provide cancer mutation data for patients at an increasingly faster rate and lower cost; however, interpreting this data remains a challenge because most observed mutations are functionally neutral. The overall goal of this study is to develop a computational method that uses protein structures to predict not only whether a mutation is non-neutral, and therefore potentially contributes to disease, but also its specific effect on the protein's biochemical function. This provides more detailed information than most previous prediction methods, which only classify mutations as neutral or non-neutral. We develop several complementary computational methods for identifying non-neutral protein mutations and their biochemical effects by determining if they occur in (1) binding sites predicted using structural models based on experimental structures of related proteins, (2) binding sites predicted by machine learning algorithms, (3) functional sites detected by mutation clusters, or (4) functional sites annotated in public databases. Finally, we find that a machine learning predictor trained on these and other features outperforms two other popular methods for detecting disease-associated mutations.

*E-mail: bordner.andrew@mayo.edu

# Introduction

Non-synonymous single nucleotide variants (nsSNVs) in the protein coding regions of the human genome, or missense variants, alter the amino acid sequence of the protein product. Some of these variants cause large phenotypic effects that lead to disease, with variants in the germline genome associated with Mendelian diseases while those in the somatic genome of cancer cells potentially contributing to tumorigenesis and cancer progression. Importantly, missense mutations are the most commonly observed protein-coding genome variant in both Mendelian diseases [6] and cancer [7]. Although rapid progress in sequencing technologies is providing mutation data faster and at lower cost, the detection of such functional variants from among neutral ones remains a significant scientific challenge. This is particularly important for cancer mutations. Because cancer cells originate through an evolutionary process of mutation, selection, and clonal expansion and typically exhibit genome instability most mutations are expected to be functionally neutral. Thus the functional, or potential driver, mutations must be distinguished from a large background of neutral, or passenger, mutations. Accurate prediction of driver mutations in cancer resequencing data has several possible medical applications, including the discovery of new drug targets, improved cancer diagnosis and prognosis, and the development of more effective personalized treatments. Additionally, the effects at the protein level of missense mutations associated with other diseases remain only partially understood.

Missense mutations can potentially affect a wide range of protein properties including thermodynamic stability, ligand (substrate, cofactor, or protein) binding, aggregation, expression level, allosteric regulation, posttranslational modification, disulfide bonds, conformational dynamics, and cellular localization. Evolutionary conservation is one of the strongest signals for detecting non-neutral mutations since purifying selection is likely operating at the site due to its functional importance and random mutations at functionally important sites are more likely to be deleterious [8]. However, conservation alone does not indicate which particular protein function or property is disrupted by a mutation at the site. In contrast, we will employ molecular modeling of the protein, which enables prediction of the specific biochemical consequences of a mutation.

## Overview

In this study, we develop three complementary approaches for identifying functional sites in protein structures and analyze the relative propensities of neutral and non-neutral missense mutations to occur within these sites. The first approach is to infer the locations of divalent metal ion, small molecule, protein-protein, and nucleic acid binding sites using homology to proteins with available X-ray structures of complexes in the Protein Data Bank (PDB) [9]. The second approach involves predicting metal ion and protein binding sites using our previously developed machine learning methods [10,11,12], as well as predicting small molecule binding sites using surface pocket detection. This approach can discover binding sites that cannot be found by homology due to lack of an X-ray structure of a homologous protein bound to the ligand of interest. Finally, the third approach is to employ spatial clustering of mutations in the three-dimensional (3D) protein structure. Because functional sites in proteins are expected to be generally localized within specific regions of the 3D molecular structure, mutations occurring within these sites, which are more likely to be non-neutral and driven by selection, will tend to cluster within these regions.

Importantly, such functional regions do not necessarily correspond to a binding site and therefore may indicate novel sites with distinct functions, for example regions that undergo large conformational changes that are crucial for the protein's biochemical function.

In order to find the most complete prediction of functional effects, we supplemented results from these methods with stability changes predicted by molecular modeling and functional site annotation from online databases. Unlike the three main approaches we investigate in this study, these have been widely employed in previous work described in the literature. First, several studies have used molecular modeling to predict stability changes and found that, as expected, many disease-associated mutations are destabilizing [13]. Second, online database annotations are included as input features for other functional mutation prediction methods such as PolyPhen-2 [14] and CHASM [15]. These annotations provide valuable information on diverse functional sites such as ligand binding sites, posttranslational modification sites, proteolytic cleavage sites, active sites, kinase activation loops, etc.. However, as shown below, our homology modeling approach detects considerably more binding sites than those annotated in the RefSeq database [16], which includes site information transferred from both the Uniprot [17] and CDD [18] databases. Finally, we trained a Random Forest classifier to discriminate functional from neutral missense mutations using input data from these computational methods as well as evolutionary conservation and additional structure and sequence properties.

### Related work

A number of previous papers have examined the potential biophysical consequences of disease mutations using structural modeling. Two computational studies by the Moult group [13,19] found that a larger fraction of Mendelian disease-associated missense mutations than neutral mutations are predicted to destabilize native protein structures. This is consistent with an analysis by Vitkup et al. [20] showing that such mutations tend to occur at sites that are inaccessible to solvent, where the potential to disrupt stabilizing physical interactions is higher. It is also consistent with the observation that most Mendelian diseases are caused by loss of function mutations, as is evident from, for example, functional assay results and the occurrence of other inactivating mutations, such as nonsense mutations, that are associated with the disease. In addition, computational analyses [21,22] have found a prevalence of Mendelian disease-associated mutations compared with neutral mutations occurring in protein-protein interfaces, where they can potentially disrupt these interactions. Finally, one previous study, by Levy et al. [23], has systematically analyzed disease-associated nsSNVs at metal ion binding sites. They found a preponderance of disease-associated nsSNVs occurred at predicted ion binding residues compared with neutral nsSNVs. Interestingly, a significant but smaller enrichment of disease-associated nsSNVs was also observed at second-shell residues, which contact the ion-ligating residues but not the ion itself. Here, we expand on these previous studies by analyzing the occurrence of disease-associated mutations from different data sets within four different classes of binding sites. In addition, we perform cluster analysis and construct a machine learning predictor for functional mutations that integrates information on predicted binding sites with additional features including functional site annotation from online databases, predicted stability effects, evolutionary conservation, and structural properties.

# Results/Discussion

## Mutation data sets

We considered three classes of mutation data sets in our analysis: Mendelian disease-associated mutations, somatic mutations in cancer, and neutral polymorphisms. Disease-associated mutation data were obtained from the Human Gene Mutation Database (HGMD, downloaded March 2012) [24]. Somatic missense mutations observed in cancer were obtained from the COSMIC database (ver. 60) [25], which collects and curates such data from multiple published studies. This data set contains both driver and passenger mutations. Furthermore, because it includes data from genome-wide sequencing studies, it likely includes a higher fraction of passenger mutations occurring in genes not implicated in tumorigenesis. Therefore, in order in generate data sets enriched in driver mutations, we selected subsets of mutations occurring in genes implicated in cancer. Two different sources of cancer genes were used. One set, from the CancerGenes website (http://cbio.mskcc.org/cancergenes) [26], contains separate lists of 495 oncogenes and 873 tumor suppressor genes. The other set, obtained from the Cancer Gene Census website (http://www.sanger.ac.uk/genetics/CGP/Census) [27], contained 487 gene implicated in tumorigenesis. Filtered cancer mutation sets were obtained by selecting COSMIC data only for mutations in the respective cancer genes. In addition, we analyzed Cancer Cell Line Project data [28], which comes from resequencing studies of 70 cancer-associated genes in different cell lines. Finally, presumably neutral polymorphisms were taken from two sources: Uniprot and the NCBI dbSNP database. The Uniprot neutral mutations were those which the online file (http://www.uniprot.org/docs/humsavar/, version 12_07) annotated as "Polymorphism". Because dbSNP data contains some non-neutral variants, it was filtered by removing those annotated as either "pathogenic" or "probable-pathogenetic", those with minor allele frequency < 5%, and those that were labeled as disease-associated mutations in the HGMD database.

## Inference of binding sites by homology

As show in **Figure 1**, only ~13% of human proteins have available high-resolution (≤ 3 Å) X-ray structures in the PDB. However, as many as ~56% of human proteins have at least one close homolog, with ≥ 30% sequence identity to a protein in an available PDB X-ray structure. Thus homology modeling can dramatically increase the coverage of the human proteome by atomic-level structural models. Furthermore, homology models can be used to infer ligand binding sites by transferring them from the template structure. Aloy et al. showed that protein-protein binding geometry is almost always similar between homologs having ≥ 30% sequence identity [29]. Although it has not been systematically studied, metal ion, small molecule, and nucleic acid binding sites are similarly observed to maintain binding geometry between close homologs.

Based on this conservation of binding geometry, we have constructed atomic-level structures of human proteins with binding sites either determined from PDB structures of the protein, when a structure of the human protein is available, or by homology modeling (see Methods for details on the procedure). Importantly, we first use BIOMT annotations to generate the biologically relevant complex before transferring binding sites. Also, because a given human protein generally has multiple homolog structures and each template structure may cover different domains of a multidomain protein or contain different ligands we create a composite model by combining homology models for distinct domains and transferring binding sites from all X-ray structures of homologs. **Figure 2** shows an example in which three different PDB structures are used to determine binding sites for the HNF4α nuclear receptor protein.

### Predicted binding sites

Machine learning methods can also be used to find metal ion, small molecule, and protein-protein binding sites, some of which may not be apparent from homology due to lack of a relevant homolog structure (see Ref. [11] for an example). We applied our previously developed methods for predicting specific divalent metal ion binding sites [10] and protein-protein binding sites in non-membrane [11] and membrane [12] proteins. These prediction methods rely on multiple signals to detect binding sites including higher evolutionary conservation and characteristic residue types at binding sites compared with the remaining protein surface. Significantly, distinctive metal ion coordinating residues allow classifiers to discriminate between different ions, for example between $Ca^{2+}$ and $Cu^{2+}$. Small molecule binding sites were simply predicted as sufficiently large (volume $\geq 100$ Å$^3$) surface pockets. Small molecules generally bind to such pockets, which remain in the unbound (apo) structure and are usually amongst the two largest surface pockets [30]. We used the ProShape program [31] to detect pockets in the X-ray protein structures.

### Occurrence of missense mutations in binding sites

**Figure 3** shows the relative occurrence frequencies of mutations from different sets in the four classes of binding sites. Detailed statistics are given in **Table S1**. One apparent trend is that disease-associated and cancer somatic mutations occur at higher frequencies in all classes of binding sites than those in the neutral mutation sets. Restricting to common polymorphisms, with MAF ≥ 5%, did not significantly affect the fraction of dbSNP mutations occurring in binding sites. Furthermore, mutations in known cancer gene sets (COSMIC CGC, COSMIC MSKCC, and Cancer Cell Lines) occurred more frequently in protein-protein binding interfaces than either Mendelian disease-associated or neutral mutations. This is especially significant since these sets likely contain neutral passenger mutations as well as functional mutations. Mutations in the Cancer Cell Lines were the most prevalent in all classes of binding sites as compared with other cancer mutation sets, or indeed any other mutation set examined in this analysis. One possible contributing factor is that the mutations were chosen from a highly selective set of only 24 cancer genes, whereas the CGC and MSKCC sets included many more genes (487 and 1368, respectively).

**Figure 4** shows the mutation occurrence frequencies in metal ion, small molecule and protein-protein binding sites predicted using machine learning, our second approach to detecting functional sites. These results manifested the same trends, albeit with different frequencies. The small molecule frequencies were probably higher than those determined by homology because the detection criteria, all surface pockets with volume 100 Å$^3$, is quite inclusive and thus likely include many false positives. On the other hand, the selected FDR cutoff of 5% yielded comparatively few metal ion and protein-protein binding residues resulting in lower mutation occurrence frequencies in these sites compared with sites inferred by homology. In particular, fewer predicted metal ion binding sites were found than in the Levy et al. study [23], probably because of the stricter requirement for homology to template structures and the multiple testing correction employed here.

## Representative examples of monogenetic disease-associated SNPs occurring in binding sites

**Mutations in hepatocyte nuclear factor 4 α (HNF4α, RefSeq:NP_000448.3) linked to mature onset diabetes of the young, type 1 (MODY1, OMIM: 125850):** HNF4α is a nuclear transcription factor that binds to DNA as a homodimer and was recently found to bind linoleic acid as an endogenous ligand [32]. The HGMD data included 40 MODY1-associated mutations, of which most (37) could be mapped to at least one X-ray structure. Furthermore, most of these mutations (25) were located in binding sites, including the homodimer (8 mutations at 7 sites), co-activator (2 mutations at 2 sites), DNA (11 mutations at 6 sites), $Zn^{2+}$ ion (3 mutations at 3 sites), and fatty acid (1 mutation) binding sites. Notably, all of the mutations at the $Zn^{2+}$ ion sites are predicted to disrupt binding according to the statistical score (see below). Because these mutations are located in a zinc finger domain responsible for recognition of specific DNA sequences such disruption is expected to also prevent proper DNA binding. As illustrated in **Figure 2**, this also provides a case illustrating the utility of combining binding sites from multiple homolog structures since no single structure contained all mutation sites. Furthermore, all homologous nuclear receptor structures had the same conserved ligand binding site, even though they contain different ligands. All of the mutations mapped to binding sites likely contribute to MODY1 by interfering with HNF4α-mediated transcription regulation. This causal link is supported by a study that found a diabetes-associated mutation in the HNF4α binding site upstream of the HNF1α gene, which is another gene mutated in MODY1 [33]. Loss of HNF4α function due to multiple causes, including reduced ligand/DNA binding or reduced transcriptional activation, are associated with MODY1 [34,35,36,37,38]. Fortuitously, a multidomain X-ray structure of HNF4α, including both the ligand and DNA binding domains, bound to DNA, fatty acid ligand, and coactivator fragment [35] was released after these calculations, permitting a retrospective analysis (PDB entry 4IQR) [35]. All mutations in binding sites agreed, including those inferred by homology from the RXRα/PPARγ structure (PDB entry 1FM9) [5]. Interestingly, the recent structure revealed that although the ligand-binding domain interface was conserved, the relative orientations of the ligand and DNA binding domains differed between these two complexes. Furthermore, several MODY1 mutations were localized to the domain-domain interface and were found to reduce transcriptional activity. Such effects could not have been predicted from previous structures since none included both domains.

**Mutations in rhodopsin (RefSeq: NP_000530.1) associated with retinitis pigmentosa (OMIM: 613731) and congenital stationary nightblindness (OMIM: 610445):** Rhodopsin is a photoreceptor required for vision in low intensity light. It is a membrane protein in the G protein coupled-receptor family. Impinging light causes isomerization of a covalently attached retinal moiety leading to a conformational change that induces G protein signaling. It is perhaps not surprising that many (19) of the disease-associated mutations from HGMD in rhodopsin occur in retinal binding residues. However a comparable number of mutations (14) occur in three of the four $Zn^{2+}$ binding sites observed in the X-ray structure (PDB entry 1HZX). Five of the mutations are in the identified high-affinity site, which was found to be critical for folding, 11-cis-retinal binding, and stability of the chromophore-receptor interaction [39]. The roles of the remaining low affinity $Zn^{2+}$ binding sites are unknown. Significantly, $Zn^{2+}$ deficiency leads to symptoms similar to retinitis pigmentosa [40] supporting the importance of $Zn^{2+}$ for proper rhodopsin signaling. A total of sixteen disease-associated mutations occur within a homodimer interface; however, eight of these are in a clearly non-biological antiparallel dimer interface, which is due to erroneous BIOMT annotation in those PDB X-ray structures. Computational methods to correct these errors should improve the identification of biological interfaces [41,42]. Also, four disease-associated mutations are in residues contacting the palmitoyl covalently attached to C322, one in the glycosylation site at N15, and another two contacting the glycan. Finally, three disease-associated mutations occur in phosphorylation sites at T339 and S342.

### Examples of functionally validated cancer driver mutations

**p53 (RefSeq: NP_001119584.1) mutations:** The p53 protein is a tumor suppressor that regulates cell cycle progression and induces apoptosis through transactivation of multiple genes. Overall, the TP53 gene encoding p53 is the most mutated gene in tumors. p53 binds DNA as a homotetramer and directly interacts with many other proteins. Due to its importance in cancer a large number of functional studies of p53 cancer mutations have been performed. We performed an analysis of the 652 p53 mutations from the COSMIC database that also have experimental information from functional assays in the IARC p53 database [43]. Here we examine a few overall features and defer more detailed analysis to a later study. First, we examined the predicted stability effects for mutations at buried residues, which are most likely to affect stability, and found that mutations resulting in loss of transactivation were predicted to be more destabilizing than those without such an effect (Wilcoxon rank sum test $p = 1.6 \times 10^{-6}$). In addition, we found that mutations occurring in either a protein-protein or DNA binding site, which were not predicted to be destabilizing ($\Delta\Delta G \leq 0$), resulted in a larger loss of transactivation activity at eight genes (WAF1, MDM2, BAX, H1433, AIP1, GADD45, NOXA, and P53R2), as measured in yeast functional assays [44], than those that did not occur in such binding sites (Wilcoxon rank sum test $p = 1.3 \times 10^{-10}$). Together, these results suggest that p53 mutations reduce transactivation through two largely independent biophysical effects, namely either loss of thermodynamic stability or disruption of protein-protein or DNA binding.

**B-cell non-Hodgkin's lymphoma:** Two studies by Pasqualucci and colleagues biochemically characterized common somatic missense mutations in the CREBBP (RefSeq: NP_004371.2) [2] and PRDM1 (RefSeq: NP_001189.2) [1] proteins found in B-cell non-Hodgkin's lymphoma samples. The first study found that many of the CREBBP mutations inactivated CREBBP acetyltransferase activity and usually in only one allele, in contrast with the usual biallelic inactivation of other tumor suppressor genes. Furthermore, they experimentally characterized the effects of selected mutations on CREBBP's ability to acetylate two of its target proteins, BCL6 and p53, and on acetyl-CoA cofactor binding. As already noted in Ref. [2], all of the common mutations in the histone acetyltransferase (HAT) domain tested occur within the acetyl-CoA cofactor binding site, whereas the two mutations not affecting acetyltransferase activity occur outside of this and any other inferred binding site. In addition, four out of the five mutants had reduced binding affinity for acetyl-CoA. These results along with stability calculations and information on whether or not the mutation was in the cofactor binding site are summarized in **Table 1**. The second study [1] examined mutations of PRDM1, a transcriptional repressor involved in B cell differentiation.

The calculation results for that protein, shown in **Table 2**, generally agree with the observed biophysical effects of the mutations. Specifically, predicted ΔΔG values for the three mutations shown to result in protein instability (p.P84R, p.P84T, and p.Y185D) were all ≥ 1.5 kcal/mol indicating significant destabilization while values for the remaining mutations were smaller. In addition, the p.C605Y mutation, which was observed to cause loss of DNA binding, was found to occur at a site where the wild-type residue coordinates a zinc ion in the zinc finger DNA-binding motif. According to the statistical scoring function, described below, this mutation is predicted to disrupt zinc ion binding, which is expected to destabilize the zinc finger domain leading to loss of DNA binding.

**Well differentiated papillary mesothelioma of the peritoneum (WDPMP):** WDPMP exome sequencing discovered the first somatic mutation of the E2F1 transcription, namely NP_005216.1:p.R166H [45]. The mutation disrupts a basic arginine residue directly interacting with DNA via hydrogen bonding to a base in the major groove based on a structure of the related E2F4 protein [46], thus interfering with specific DNA sequence recognition.

## More binding sites discovered by homology than by online annotations

We have included binding site annotations from the Uniprot and CDD databases (taken from RefSeq) along with binding sites identified by homology. Therefore it is interesting to compare these two sources of binding site information, online databases and inference by homology, in terms of their completeness and complementarity. **Table 3** shows the number of binding sites in each source as well as the number of sites contained in one source but not the other. There are roughly twice as many proteins with sites inferred by homology compared with those from the database annotations for each ligand type. Consistent with this, there are considerably more proteins that have a binding site inferred by homology but not included in the databases than the converse. However, since there are still a substantial number of binding sites from online annotations that were not inferred by homology, likely because of no suitable template structure of the complex, it is advantageous to include the sites from online databases along with structurally inferred sites.

## Spatial clustering of mutations

Many functional sites in proteins are contained within well-defined local regions of the 3D structure. These include enzyme active sites, ligand binding sites, and kinase activation loops. Because mutations in these sites have the potential to directly disrupt the corresponding protein function one expects that disease-associated mutations may be enriched in them. Motivated by this rationale, we have attempted to detect novel functional sites in proteins as statistical significant spatial clusters of mutations within the 3D protein structure.

Previously, Ye et al. [47] used a statistical measure to discover non-random clusters of somatic mutations in the linear amino acid sequences from cancer samples. They were able to detect known mutation hotspots in oncogenes, including KRAS, BRAF, RAS genes, PI3K, ERBBB2/Her2, and CTNN1B1/β-catenin. Furthermore, they discovered fewer mutation hotspots in tumor suppressors than in oncogenes. One advantage of our approach of clustering mutations in the 3D structure rather than the 1D sequence is that it can detect clusters that are spatially localized in the protein structure but include multiple discontinuous segments in the amino acid sequence. This is expected to allow the detection of more statistically significant mutation clusters.

A possible confounding source of mutation clusters in cancer could be due to other factors unrelated to functional protein sites, such as DNA recombination hotspots or artifacts of DNA repair. For example, the study by Roberts et al. [48] found clusters of cancer mutations within the genome sequence that appear to have resulted from multiple mutations within single-strand DNA segments near repaired double strand breakpoints or chromosome rearrangement breakpoints. Another study by Amos found clustering of human SNPs in the genome sequence, most of which are likely neutral and occur near heterozygous sites [49]. By definition, clusters resulting from these alternative mechanisms will be localized in the linear DNA sequence and not span multiple contiguous segments in the 3D protein structure as clusters of functional mutations are expected to do.

One previous study, by Stehr et al. [50], has investigated the spatial clustering of cancer mutations by calculating the statistic $(1/N)\sum_{i,j} 1/d_{i,j}$, in which the $N$ $d_{i,j}$ values are distances between residue side chain centroids within a single domain. They found statistically significant clustering of SNPs in all data sets examined, including cancer, monogenetic disease, and neutral mutations. Whereas Stehr et al. demonstrated the overall tendency of mutations to cluster, here our goal is to actually detect the individual mutation clusters using a Bayesian clustering algorithm that fits the number, center, and shape of each cluster. Based on the hypothesis that these clusters represent functionally important regions of the protein structure, they can then be used to identify non-neutral mutations. In addition, Wagner [51] used spatial clustering of non-synonymous variants with the different goal of detecting positive selection and applied it to human-chimpanzee ortholog pairs.

We performed spatial clustering using the Bayesian algorithm (details in the Methods section) applied to the HGMD, COSMIC, and dbSNP mutation data sets. Next, we compared the number of oncogene and tumor suppressor proteins with at least one mutation cluster for both cancer and neutral mutations taken from the COSMIC and dbSNP databases, respectively. Statistical analysis of the results, shown in **Table 4**, indicated that clusters of cancer mutations occurred more frequently in oncogenes than tumor suppressors, in agreement with the analysis of Ye et al. [47]. One possible explanation for this difference is that tumor suppressor mutations are mostly inactivating and therefore can occur at multiple sites throughout the protein whereas oncogene mutations are activating mutations that are likely localized in regions important for regulation. In contrast, no such significant difference was observed for neutral polymorphisms. Also, the overall fraction of cancer-associated proteins (oncogenes and tumor suppressors) with clusters compared with all proteins containing clusters was higher for the COSMIC cancer mutation data. This suggests that the discovered mutation clusters are likely to correspond to functional regions of the proteins that are implicated in tumorigenesis and cancer progression. Furthermore, the number of neutral mutation clusters is proportional to the number of genes in the set and occur in less than 5% of the cancer (tumor suppressor and oncogene) proteins, suggesting that these clusters predominantly occur by chance and therefore do not represent functional regions.

In addition, we compared these results with clusters discovered by applying the same Bayesian clustering algorithm with identical FDR cutoff to the amino acid sequence. The comparison results, shown in **Figure 5**, clearly show that 3D clustering was able to detect more statistically significant clusters and also found the majority of clusters detected by 1D sequence clustering. Thus 3D clustering is a more sensitive detection method for proteins whose structures can be modeled.

We also examined the number of proteins with discovered spatial clusters that corresponded with binding sites inferred by homology. Binding sites for all four ligand classes included in **Table S1** were considered and only cases where the percentage of binding site residues in the overlap was ≥ 50% were included. Only 15/110 (13.6%) of the proteins in the COSMIC set and 17/96 (17.7%) of the proteins in the HGMD set contained clusters that significantly overlapped with a binding site (see **Figure 6**). The overlap with binding sites predicted by machine learning was only slightly higher with 20/110 (18.2%) of the proteins in the COSMIC set and 21/96 (21.9%) of the proteins in the HGMD set containing a mutation cluster with significant overlap. Additionally, none of the clusters of dbSNP neutral mutations overlapped with a binding site predicted either by homology or machine learning, suggesting that the few clusters found in that set are likely to be false positives. Importantly, these results demonstrate that spatial clustering of mutations can be used to discover potentially novel functional sites in addition to binding sites inferred by homology.

### Evolutionary conservation
Although functional mutations have been previously been shown to occur predominantly at evolutionarily conserved sites compared with neutral mutations, we attempted to confirm this with the data sets used in this study. The column entropy calculated from a multiple sequence alignment was used as a simple measure of conservation with lower values reflecting higher conservation (see Methods for details). We found that mutations from data sets enriched in functional mutations (HGMD, COSMIC, and COSMIC Cell Lines) all had significantly higher evolutionary conservation than neutral mutations from the dbSNP data set (Wilcoxon rank-sum test, $p < 2.2 \times 10^{-16}$). This trend is in agreement with previous studies, which found that disease-associated mutations occur more frequently at conserved sites [8,52]. Indeed, evolutionary conservation is arguably the strongest signal for differentiating disease from neutral mutations and is used as an input variable in virtually all disease mutation prediction methods.

### Destabilizing mutations

We also compared the fractions of mutations in each set that were predicted to significantly destabilize the protein by at least 2 kcal/mol. These results are plotted in **Figure 7** with numeric values given in **Table S1**. As expected, neutral mutations from dbSNP and Uniprot had the lowest prevalence of destabilizing mutations. Interestingly Mendelian disease mutations from HGMD had almost twice the fraction of destabilizing mutations compared with the cancer mutations. One possible explanation is that destabilizing mutations result in loss of function as do almost all Mendelian disease mutations while many cancer mutations, particularly those in oncogenes, result in gain of function and therefore are unlikely to cause protein unfolding. The large proportion of Mendelian disease mutations predicted to cause thermodynamic instability agrees qualitatively with the results of Refs. [13,19], although we find a lower absolute proportion, due in part to our conservative (high) ΔΔG cutoff of 2 kcal/mol.

### Machine learning prediction of functional versus neutral missense variants

We also trained a Random Forest classifier [53] to detect non-neutral mutations using information on whether the mutation occurred in a predicted binding site along with other features including online database annotations and additional structural properties. **Table S2** provides a list of all input features used for the prediction algorithm. The Random Forest machine learning method is optimal for this prediction task because it is insensitive to feature normalization, can handle both categorical and continuous values, only has one adjustable parameter, and is relatively robust to overfitting.

In order to evaluate prediction accuracy, we performed 10-fold cross-validation on a set consisting of the HGMD mutations and an equal number of mutations from dbSNP. We chose the optimal number of variables/tree parameter to be 12, although there was little variation in the AUC performance measure (< 0.01) over the the wide range of 8-20 variables. This insensitivity to the model parameter is typical for Random Forests. The Random Forest predictor had an AUC of 0.80.

Next, we compared our prediction approach against the widely used SIFT [54,55] and PolyPhen-2 [14] methods. In order to obtain accurate estimates of prediction performance, we started with HGMD (dbSNP) as functional (neutral) mutation data sets and then removed all mutations included in the Uniprot set, which were used for training the PolyPhen-2 predictor. We also included only mutations that map to at least one homolog with a PDB structure. All remaining HGMD mutations not included in the test set were used in the training set for our method. Finally, we randomly selected equal numbers of mutations from the much larger neutral mutation set to create a balanced test set of 19038 mutations and put all remaining neutral mutations into the training set. Thus all three methods were evaluated on the same balanced test set and our method was trained on independent mutation data, as the others were presumably too. Probably because of the relatively small training set, our Random Forest predictor performed slightly better with an unbalanced training set containing more neutral mutations from dbSNP than functional ones despite the fact that the test set was balanced. We therefore used the complete unbalanced training set, which had an approximate 17:1 ratio of neutral to functional mutations. The results of this comparison showed that our Random Forest variant classification method yielded a higher AUC value of 0.802 compared with 0.658 for SIFT and 0.736 for PolyPhen-2. **Figure 8** shows the relevant portion of the ROC curve for low false positive values. We also found that the AUC value for our method decreased to 0.789 when homology-inferred binding site data was omitted and decreased further to 0.635 when all data calculated from the homology model structures was removed, i.e. keeping only evolutionary conservation, RefSeq annotations, dbPTM information, and wild-type/mutation residue type information. These results demonstrate the importance of including structure-based properties for accurately predicting functional missense mutations.

### Mutations in metal ion binding sites are likely to be disruptive

The propensity of disease-associated mutations to occur in binding sites brings up the important question of whether they disrupt or promote ligand binding. Based on biophysical studies, random missense mutations have been found to overwhelmingly reduce protein stability [56] and are also expected to generally reduce binding affinity, given the multiple favorable atomic interactions required to achieve specific ligand binding. Furthermore, higher sequence conservation at binding sites due to purifying selection [11,57] suggests that most residue substitutions disrupt these interactions. However, evolutionary selection processes leading to cancer or to Mendelian diseases through heterozygote advantage could potentially cause mutations that increase ligand binding. Molecular modeling is one promising approach to predict the effect of a mutation on ligand binding, which we will pursue in later studies. Here, we use a simpler statistical approach to infer how a mutation affects binding. We will examine divalent metal ion binding sites, which because of their small size and the characteristic binding residues and geometry of each ion [58,59] are particularly amenable to this approach.

We first fit a simple multinomial probabilistic model for the occurrence of each residue type in binding sites for each of the six different divalent metal ions considered above, again concentrating only on residues directly contacting the ion. Next, we examined the relative number of cases in which the expected occurrence probability, or multinomial p-value, was lesser or greater after each residue mutation. Since the multinomial parameters are estimated as the frequency of each residue type in metal ion binding sites and the mutation only involves one site, this is equivalent to comparing the occurrence frequencies of the mutant and wild-type residue types. In order to get an independent estimate of binding site residue frequencies we collected non-redundant sets of proteins (using a 25% sequence identity cutoff) with available X-ray structures in the PDB that have the ion of interest bound. We then calculated the frequencies of each of the twenty residue types in binding sites for each ion type. Notably, as is apparent from **Figure 1**, these proteins are predominantly from non-human organisms and thus presumably reflect the specific physiochemical interaction propensities of each metal ion rather than any human-specific binding motifs. These ion-coordinating residue frequencies are plotted in **Figure S1** and given in **Table S3**.

Next, we applied the statistical scoring function to disease-associated and neutral mutations, from HGMD and dbSNP respectively, that occur in metal ion binding sites inferred by homology. The results of this analysis are shown in **Table 5**. Significantly, disease-associated mutations in all six types of metal ion binding sites were predicted to disrupt binding. Interestingly, presumed neutral mutations from dbSNP were similarly predicted to disrupt binding for four of the ion types with the most mutation data ($Ca^{2+}$, $Mg^{2+}$, $Mn^{2+}$, and $Zn^{2+}$). One contributing factor could be that some dbSNP mutations are actually associated with disease but are not annotated as such. Another possibility is that loss of ion binding in some cases may be near neutral, mildly deleterious, or even confer a heterozygous advantage and therefore occur at moderate levels in the population [60]. This hypothesis that many polymorphisms have significant biochemical effects is supported by the experimental study of Allali-Hassani et al. [61], in which they measured the effects of a small set of non-synonymous polymorphisms in several enzymes that are not yet linked to disease and found that a significant fraction altered thermodynamic stability and catalytic activity, suggesting that many presumably neutral SNPs may in fact alter normal human physiology or even contribute to disease. Finally, a disruptive mutation could be in fact be rare but not labeled as such since allele frequency information is only available for a small fraction of dbSNP variants. Many such rare variants are predicted to have significant functional impacts [62].

## Conclusions

In this study, we have performed a proteome-wide analysis of the relative occurrence frequencies of missense variants from different data sets in ligand binding sites and found that Mendelian disease-associated variants and cancer somatic variants are more prevalent in all four classes of binding sites than neutral polymorphisms. We also found that disease-associated mutations are much more likely to form statistically significant spatial clusters in the protein structure than neutral mutations. Therefore, mutations occurring within these two types of sites are more likely to be associated with disease, presumably because they directly affect the corresponding ligand binding or protein function mediated by residues within the clusters. We included predicted binding site information along with functional site annotations from online databases, predicted stability effects, and evolutionary conservation to train a machine learning classifier to discriminate functional from neutral mutations. The resulting classifier was found to outperform the SIFT and PolyPhen2 methods when applied to an independent mutation test set.

Significantly, we have also shown that this structure-based approach can be applied to a significant fraction (56%) of human proteins for which template structures of close homologs are available for reliable homology modeling. This coverage of the human proteome is expected to rapidly increase in the future due to high-throughput experimental methods employed by structural genomics projects. In fact, many such projects have the primary goal of solving structures of human proteins or novel folds that are likely to increase the coverage of homology models [72].

A key advantage of using protein complex structures to analyze missense mutations is that it predicts the possible biochemical consequences of each mutation. These predictions can subsequently be validated in experiments and thereby provide valuable mechanistic information that can inform efforts to develop new disease treatments. For example, biophysical measurements can be used to test whether mutations occurring within binding sites perturb binding affinity. Although mutation clusters provide more indirect clues on functional effects than binding sites, they still generate experimentally testable hypotheses. In particular, such localized clustering in the 3D protein structure suggests that mutations within each cluster affect the same protein function. For example, mutations within a cluster containing known kinase activating mutations are more likely to lead to similar activation.

### Future directions

There are several directions for future work extending the structural modeling approach described in this paper. First, because homology models of complexes are available, it would be interesting to predict the change in binding affinity due to each mutation using molecular modeling. This would allow each mutation within a binding interface to be ranked by its predicted change on binding affinity. Second, biological pathway analysis could be applied to infer the cellular effects of biochemical perturbations caused by disease-associated mutations. Another, more specific, improvement would be to use computational methods to predict the correct biological complex present in the X-ray crystal structure, which is crucial for generating accurate models of human protein complexes. Previously, we found a significant fraction of structures with errors in author-provided or computationally predicted information contained in the BIOMT records of PDB files and developed a prediction method that could be used for this purpose [41]. Finally, while the 30% sequence identity cutoff is appropriate for inferring protein-protein interactions from structures of homologs according to Aloy et al. [29], a lower cutoff may be optimal for interactions with other types of ligands, due to higher evolutionarily conservation; however, further study is required. If this is the case, it would increase the total number of binding sites that can be identified in human proteins using homology modeling.

## Methods

### Mapping nsSNVs to protein structures

First, we generated pairwise sequence alignments between each human protein in RefSeq (updated Oct. 15, 2012) and proteins with X-ray structures in the Protein Data Bank (PDB) [9]. In order to include only reliable structures, only those with a resolution of 3.0 Å or better were considered. Amino acid sequences were aligned by searching all PDB protein sequences using PSI-BLAST [63] with a sequence profile generated using two iterations of PSI-BLAST applied to the nr database and using an E-value cutoff of $10^{-3}$. Only alignments with sequence identity ≥ 30% and ≥ 80% coverage of the PDB protein sequence were retained. Importantly, a significant fraction, 19,250/34,677 (56%), of the human proteome was covered by either a structure of the human protein or a structure of a homolog for use in comparative modeling were available in the PDB. The inclusion of structures for homologs greatly expanded the coverage since, by contrast, PDB structures were available for only 4497 (13%) of the human proteins (**Figure 1**) at the time of this analysis.

Functional site information was obtained from RefSeq "Site" annotations [16]. These annotations are propagated from both the NCBI Conserved Domain Database (CDD) [18,64] and UniProt [17] databases.

### Ligand classes

We considered binding sites in the protein structures for four different ligand classes: divalent metal ions ($Ca^{2+}$, $Cu^{2+}$, $Fe^{2+}$, $Mg^{2+}$, $Mn^{2+}$, and $Zn^{2+}$), small molecules, nucleic acids (DNA and RNA), and other proteins. Protein binding sites were determined by first generating the biological complex according to the BIOMT record in the PDB file. We chose an inclusive definition of small molecule ligands since many are not biological ligands but rather analogs or drugs that bind to the same site on the protein as the native ligands. All PDB heterocompounds were included in the set of small molecules except non-specific small molecules, which are often added to facilitate crystallization, and other metal ions. These include most molecules in the Russell lab set (http://www.russelllab.org/wiki/index.php/Non-specific_ligand-binding) as well as some additional ones. A complete list of these excluded heterocompounds is given in **Table S4**. Binding site residues were required to contact the ligand, with contacts defined by non-hydrogen atom separation ≤ 4 Å.

### Machine learning prediction of binding sites

As part of our second approach, divalent metal ion and protein binding sites in each human protein structure were predicted using our previously developed machine learning methods [10,11,12], which are based on Random Forest classifiers [53]. Separate predictors were used for each of six common divalent metal ions ($Ca^{2+}$, $Cu^{2+}$, $Fe^{2+}$, $Mg^{2+}$, $Mn^{2+}$, and $Zn^{2+}$). For prediction of protein-protein binding sites, each protein was first classified as either a non-membrane or integral membrane protein according to the PDBTM database [65] and the corresponding machine learning classifier was then applied to the human protein. The raw output score from the Random Forest classifiers used for predicting binding sites represents the fraction of classification trees voting for the positive class (binding site residue), which varies from 0.0 to 1.0. Since these score do not have a direct interpretation in terms of prediction confidence we converted them into probabilities, which do quantify confidence. This was done by estimating the empirical probability density function $p_-(S)$ for labeled negative (non-binding site) data using kernel density estimation via the "density" function in R [66]. The p-value corresponding to a score S was then calculated as $1 - D_-(S)$, in which $D_-(S)$ is the cumulative probability distribution corresponding to $p_-(S)$. Finally, these p-values were corrected for multiple testing using the Benjamini-Hochberg procedure [67] and sites with scores below the 20% FDR cutoff were predicted to be in the corresponding type of binding site.

### Spatial clustering of mutations

In our third approach, we attempted to detect statistically significant spatial clustering of missense mutations mapped to protein structures, as described in the Methods section. The different classes of mutations were separately analyzed. Also, only human proteins with at least ten mutations from the data set of interest were considered in order to have sufficient data to reliably infer clusters. Because the number of clusters present in each structure is unknown *a priori*, we employed the model-based Bayesian clustering procedure implemented in the MCLUST package [68] in R [69], which determines the optimal clusters as well as their number. Briefly, this clustering procedure involves fitting a Gaussian mixture model with a full covariance matrix (*i.e.* the most general 3D Gaussian function) and the number of components varying from 1 to $N_{max}$, which we chose to be 10. The optimal model is the one with the minimum Bayes Information Criterion (BIC) [70]. The BIC statistic penalizes models with a large number of parameters and thereby reduces overfitting; otherwise, simply selecting models based on maximum likelihood would always result in the model with the largest number of parameters (i.e. number of clusters) as the best one. Finally, each data point is assigned to the most probable component in the mixture model. The statistical significance of each cluster was assessed assuming a null model with a uniform distribution of mutations across sites in the protein structure.

### Statistical tests

Unless stated otherwise, statistical hypotheses were tested using the Fisher exact test at the 5% significance level. In cases of multiple hypothesis testing, the false discovery rate was limited to 5% using the Benjamini-Hochberg procedure [67].

### Predicting stability changes of mutations

We calculated the predicted change in folding free energy due to a mutation using our previously published method [71] applied to a homology model of the human protein. The homology model was generated using a simple procedure implemented in the ICM molecular modeling program (Version 3.7, Molsoft LLC), as follows. First an extended peptide model with ideal covalent geometry (bond lengths and angles) was generated from the amino acid sequence. Next, harmonic restraints were defined between corresponding atoms in the model and the template PDB structures with residue correspondences based on a pairwise sequence alignment. These restraints were used to align the model to the template by local optimization of the restraint energy in torsion angle space. Finally the model was refined to remove any steric clashes or other energetically unfavorable interactions by iteratively minimizing the sum of the restraint and physical (ECEPP/3) energy while gradually reducing the harmonic restraint weight at the start of each iteration.

### Acknowledgements

This work was funded by an Arizona State University/Mayo Clinic Seed Grant.

## Figures

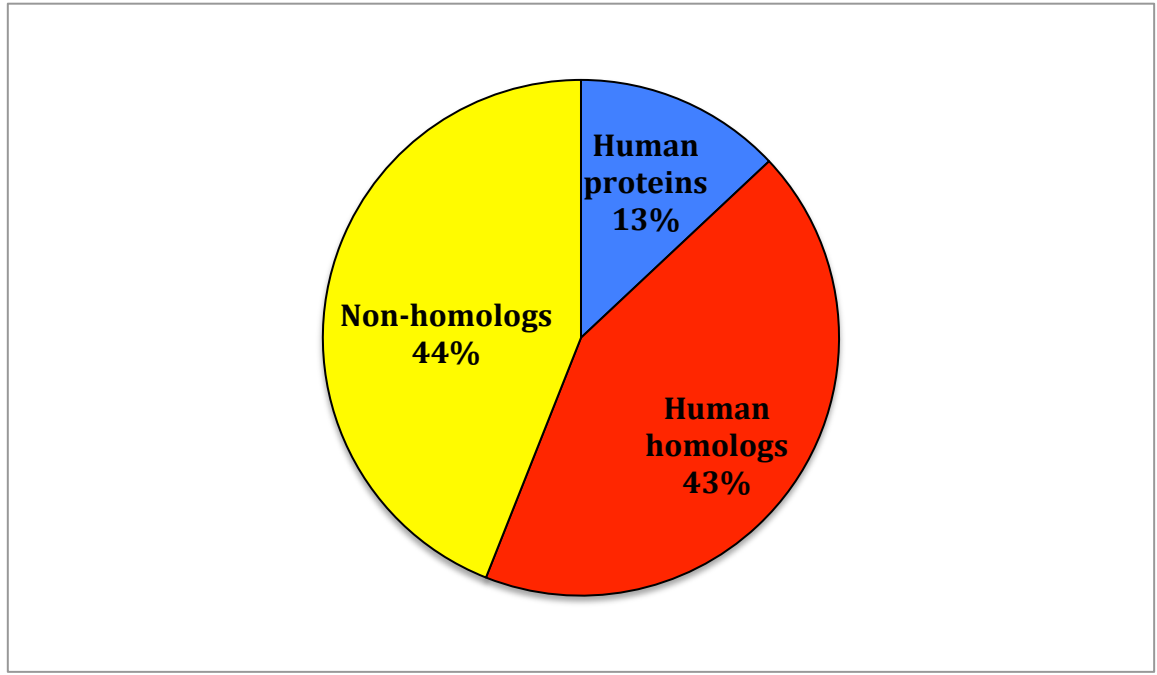

**Figure 1: Percentages of human proteins, human protein homologs, and non-homologs with high-resolution X-ray structures (≤ 3 Å) in the Protein Data Bank.** All homologs are required to have a sequence identity ≥ 30% and alignment coverage ≥ 80% with at least one human protein. These proportions show that inferring functional sites in human protein structures by homology dramatically increases the coverage of the human proteome by atomic models.

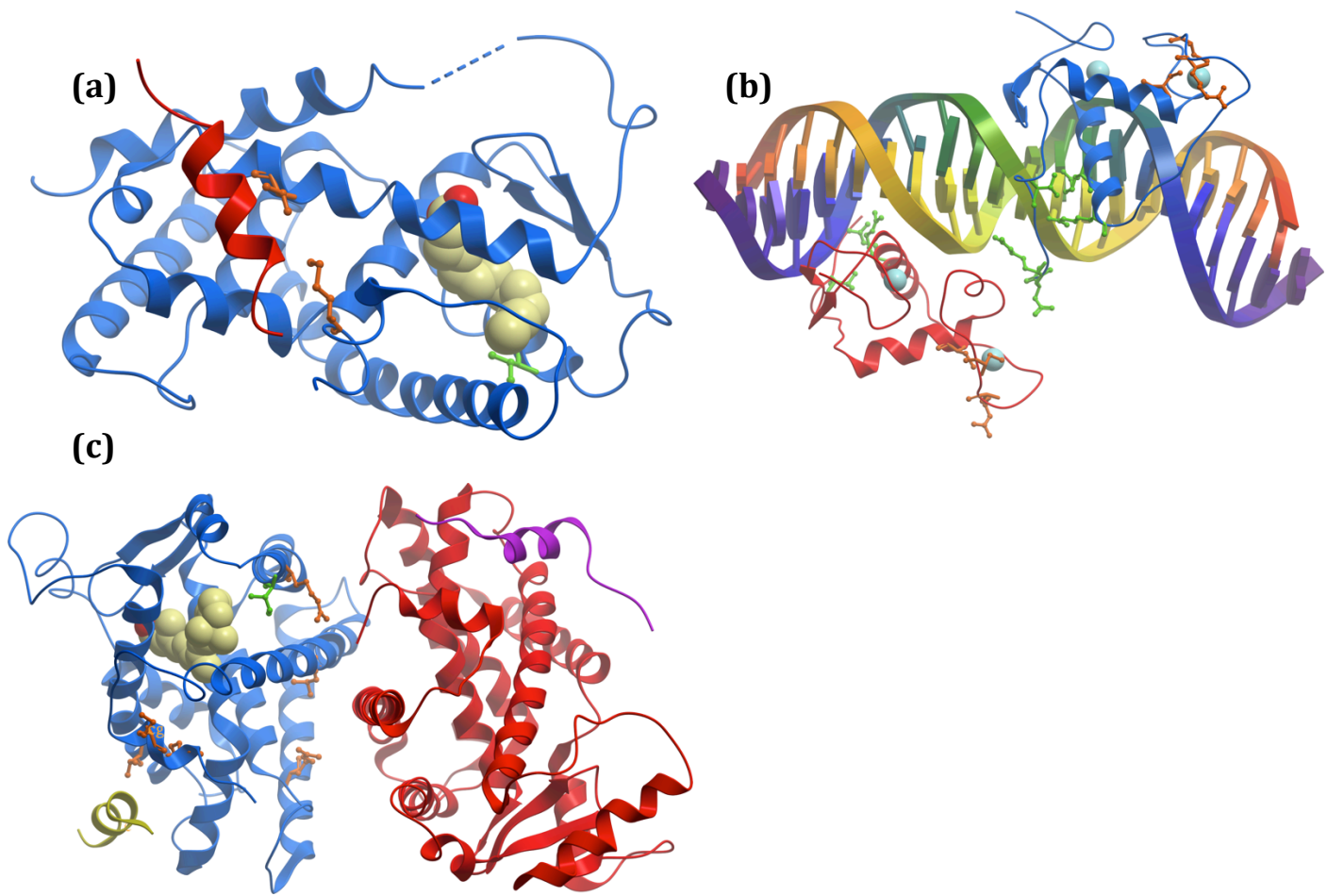

**Figure 2: Mutations in human HNF4α nuclear receptor associated with MODY1 adult onset diabetes map to distinct binding sites in three different X-ray structures of complexes.** Figure (a) shows the ligand binding domain of human HNF4α bound to a fatty acid ligand and SRC-1 coactivator fragment (PDB entry 1PZL [3]). One mutation, NP_000448.3:p.V264M (shown in green), contacts the fatty acid ligand while two others, (shown in orange) contact the coactivator α-helix. Figure (b) shows a structure of the DNA binding domain homodimer bound to DNA and four $Zn^{2+}$ ions (shown in light blue) (PDB entry 3CBB [4]). Residues at MODY1 mutation sites contacting DNA (green) and $Zn^{2+}$ ions (orange) are also shown. Figure (c) shows the X-ray structure of a heterodimeric complex between the ligand binding domains of RXRα and PPARγ, which are both HNF4α nuclear receptor homologs (PDB entry 1FM9 [5]). The structure also contains bound retinoic acid and SRC-1 coactivator fragments. The RXRα/PPARγ heterodimer can be used as a template to model the HNF4α ligand binding domain homodimer. Importantly, both the ligand and coactivator α-helix binding sites are conserved since they coincide with those in the corresponding HNF4α ligand binding domain structure shown in (a). MODY1 mutation sites contacting the protein partner and coactivator (orange) and the ligand (green) are also shown. This example demonstrates the advantage of aggregating binding sites inferred from multiple X-ray structures since no single structure contains all relevant HNF4α mutation sites.

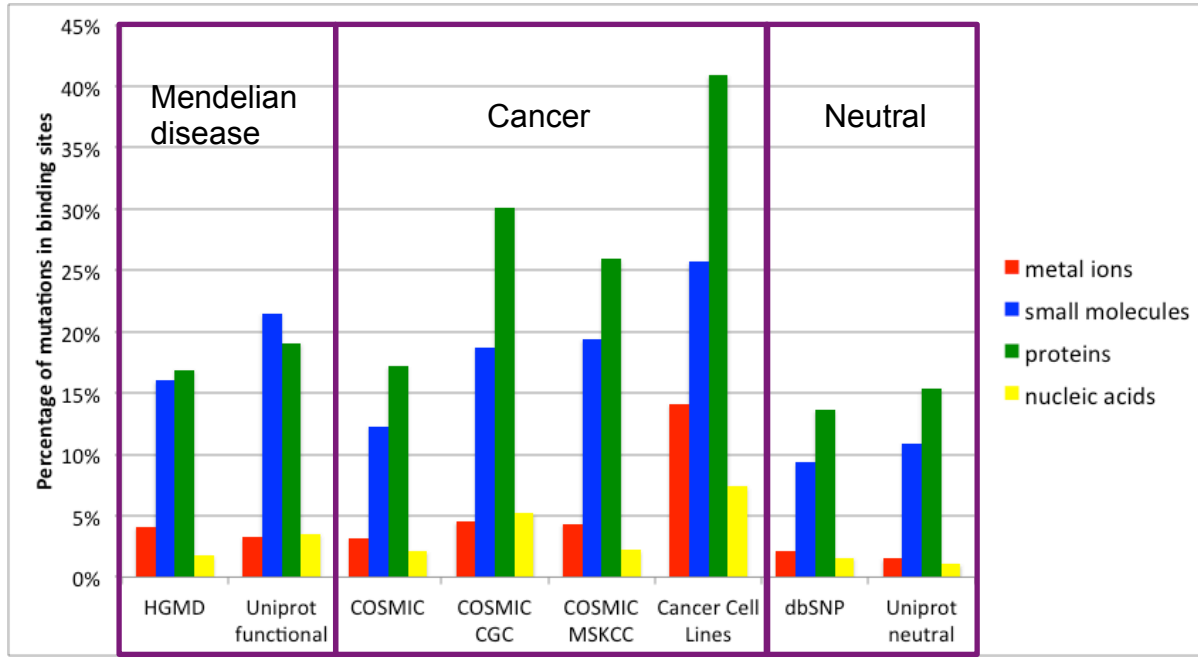

**Figure 3: Percentage of missense variants from different data sets occurring within each of the four classes of binding sites inferred by homology.** All values are listed in **Table S1**. Boxes indicating grouping of mutation data sets as Mendelian disease mutations, cancer somatic variants, and neutral polymorphisms, from left to right.

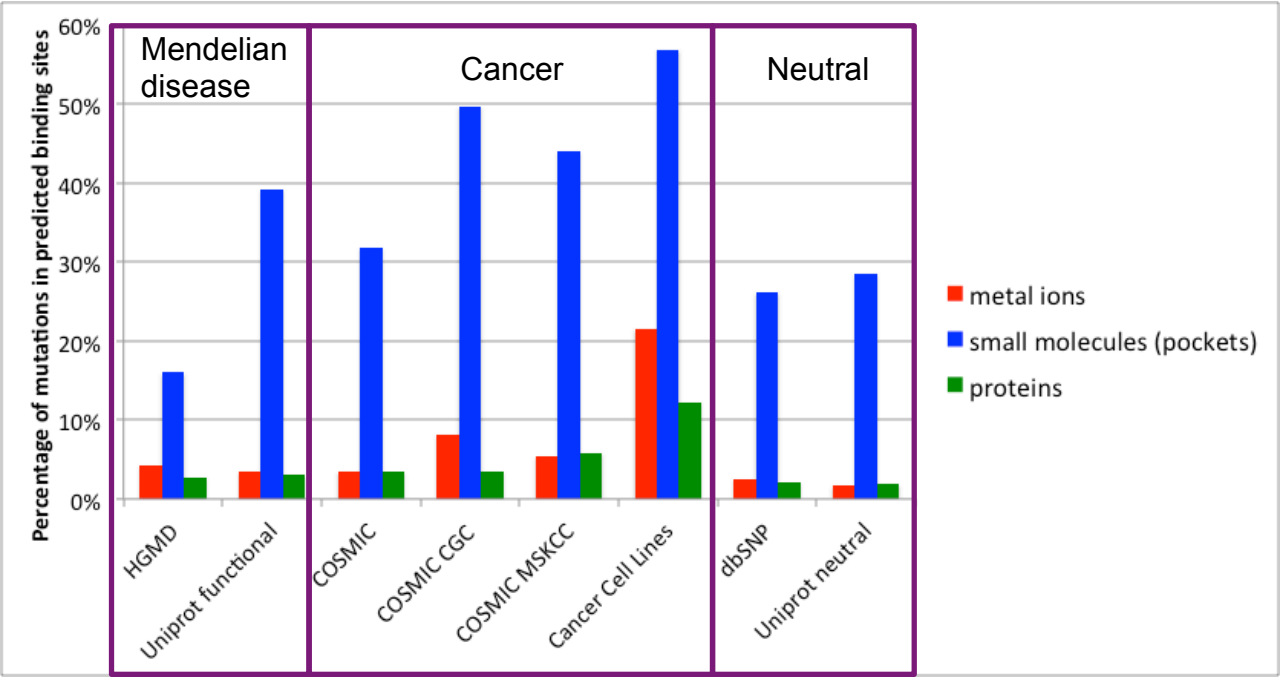

**Figure 4: Percentage of missense variants from different data sets occurring metal ion and protein binding sites predicted by machine learning or surface pockets, which are predicted as small molecule binding sites.** All values are listed in **Table S2**.

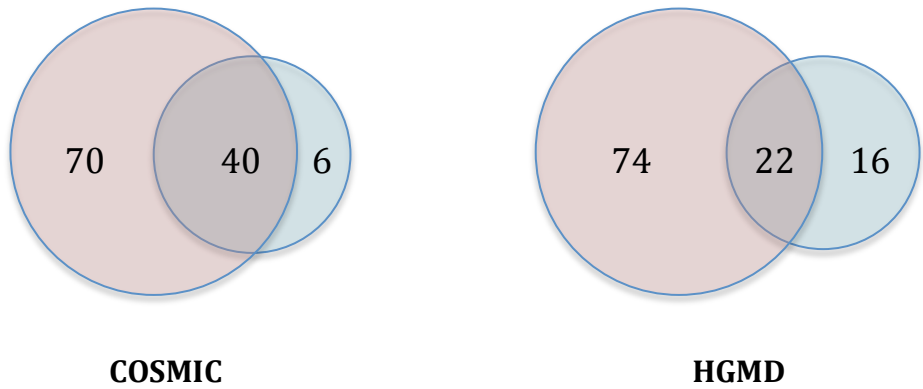

COSMIC                              HGMD

**Figure 5: Number of proteins with at least one mutation cluster detected using clustering in the 3D protein structure (red) compared with clustering in the linear amino acid sequence (blue).** These results show that spatial clustering is able to detect more potential functional sites than linear clustering.

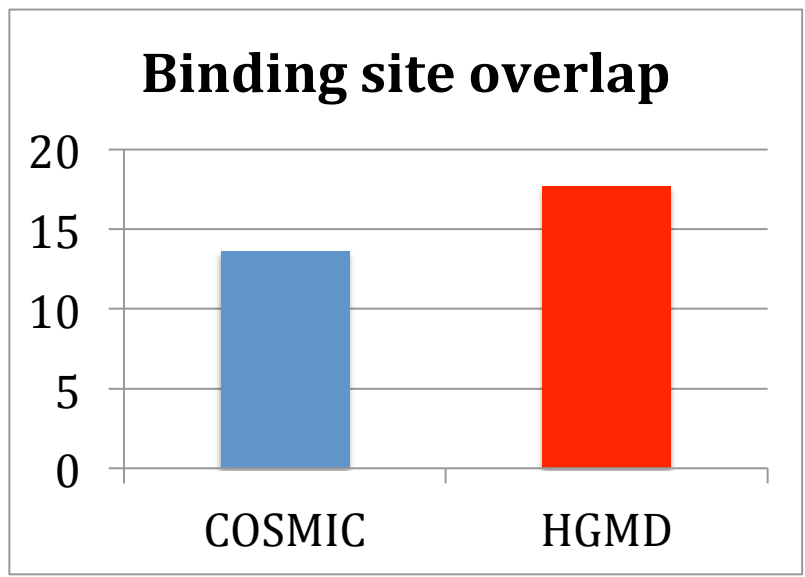

**Figure 6: Fraction of mutation clusters discovered in the COSMIC and HGMD data sets that overlap at least 50% with a binding site inferred by homology.**

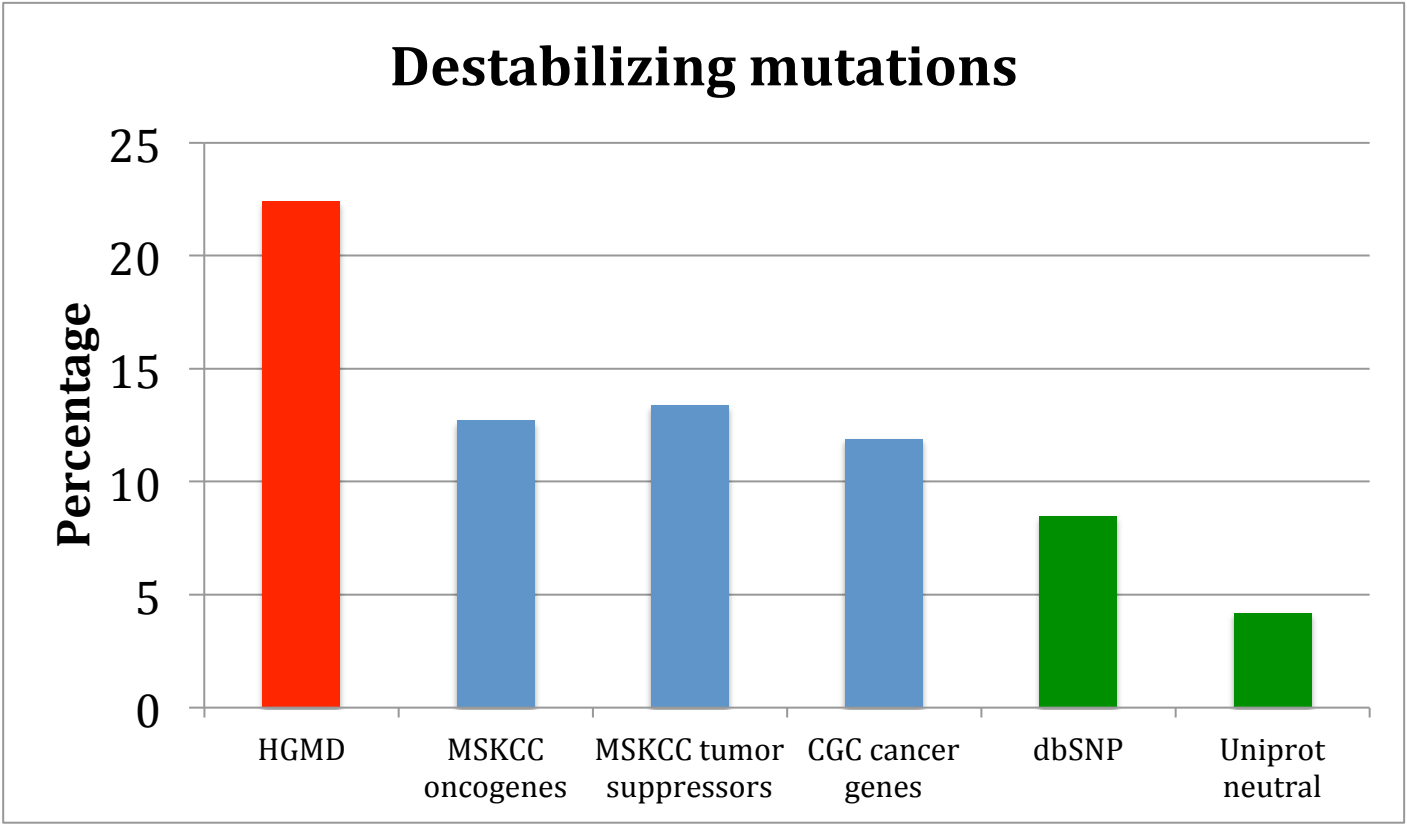

**Figure 7**: Percentage of mutations predicted to be significantly destabilizing, with ΔΔG > 2 kcal/mol, for each mutation data set. Results for Mendelian disease mutation (HGMD) are shown in red, those for cancer mutations (different subsets of COSMIC) are shown in blue, and those for neutral mutations are shown in green.

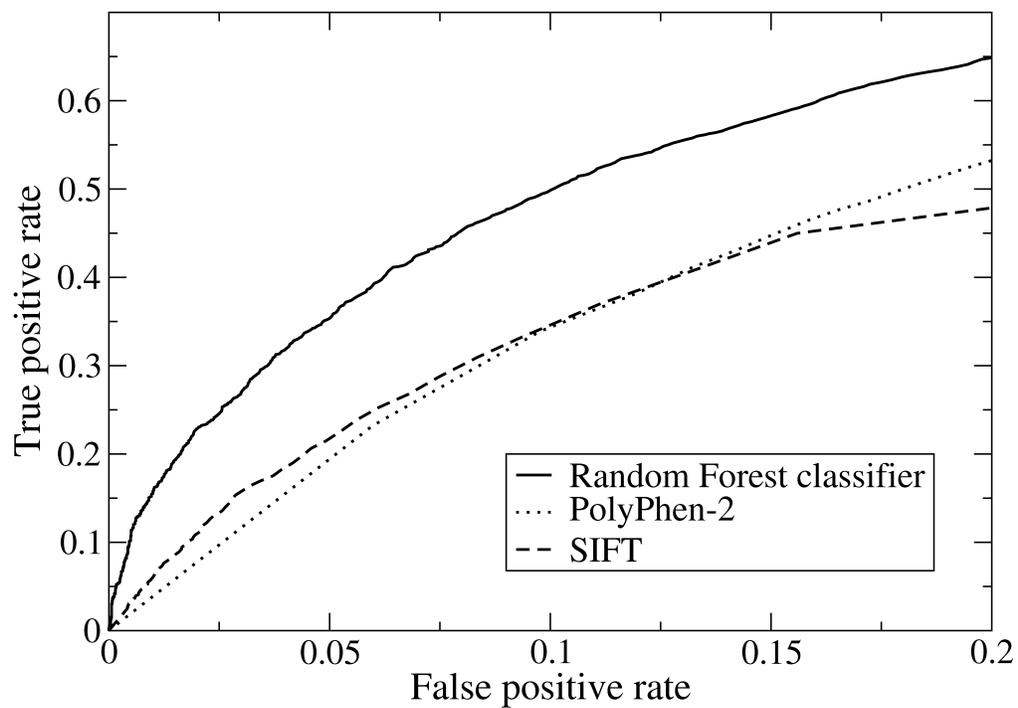

**Figure 8:** ROC curves produced by applying the Random Forest method described in this paper, PolyPhen-2, and SIFT to the same independent mutation test set. Only the portion relevant for practical use, corresponding to low false positive rates, is shown.

## Tables

Table 1: Summary of experimental results for the biochemical effects of common somatic CREBBP mutations in B-cell non-Hodgkin's lymphoma and follicular lymphoma from Pasqualucci *et al.* 2011 [2] and information on predicted stability changes and whether the mutation occurred in the cofactor/substrate binding site.

| Mutation (DNA) | Mutation (protein) | Loss of acetylation activity? | Reduced acetyl-CoA binding* | Binding site | $\Delta\Delta G$ (kcal/mol) |
|---|---|---|---|---|---|
| c.3362C>T | p.P1053L | No | - | None | 0.094 |
| c.3441G>T | p.Q1079H | No | - | None | 0.043 |
| c.4226G>A | p.R1341Q | No | - | None | 1.8 |
| c.4509T>A | p.D1435E | Yes | No | Contacts acetyl-CoA | 0.51 |
| c.4541G>A | p.R1446H | Yes | - | Contacts acetyl-CoA | 1.4 |
| c.4541G>T | p.R1446L | Yes | - | Contacts acetyl-CoA | -0.63 |
| c.4553A>G | p.Y1450C | Yes | Yes | Contacts acetyl-CoA | 1.9 |
| c.4648T>C | p.Y1482H | Yes | Yes | Contacts acetyl-CoA and lysine | 1.2 |
| c.4663C>T | p.H1487Y | Yes | Yes | Pocket near acetyl-CoA | 0.28 |
| c.4711T>G | p.Y1503D | Yes | Yes | Contacts acetyl-CoA | 2.6 |
| c.4712A>T | p.Y1503F | Yes | - | Contacts acetyl-CoA | 1.6 |

DNA and protein mutations refer to NM_004380.2 and NP_004371.2, respectively.
*Missing value indicates that the mutation was not tested.

**Table 2: Physical effects of missense mutations in PRDM1 found in B-cell non-Hodgkin's lymphoma samples reported in Mandelbaum *et al.* 2010 [1].** The mutations observed to destabilize the protein were predicted to be either marginally (P84T) or strongly (P84R and Y185D) destabilized while the remaining mutation (C605Y) was not predicted to be destabilizing but rather coordinated a $Zn^{2+}$ ion in a zinc finger domain. Based on the probabilistic model described in the Results section, the C→Y mutation is predicted to significantly disrupt $Zn^{2+}$ ion binding, which would lead to the observed loss of DNA binding by destabilizing the zinc finger domain.

| Mutation (DNA) | Mutation (protein) | Effect | Binding site | ΔΔG (kcal/mol) |
|---|---|---|---|---|
| c.485C>G | p.P84R | Protein instability | None | 3.5 |
| c.484C>A | p.P84T | Protein instability | None | 1.5 |
| c.787T>G | p.Y185D | Protein instability | None | 3.8 |
| c.2048G>A | p.C605Y | Loss of DNA binding | $Zn^{2+}$ in zinc finger motif | -0.92 |
| c.648C>G | p.I138M | None observed | None | 0.90 |
| c.2378C>T | p.A715V | None observed | None | 0.16 |

DNA and protein mutations refer to NP_001189.2 and NM_001198.3, respectively.

**Table 3: Comparison of the number of human proteins with at least one binding site annotated in RefSeq with those inferred by homology for each ligand type.** The number of proteins with binding sites determined by homology is consistently higher than the number with RefSeq annotations for all ligand types. This implies that the homology modeling method described in this study can be used to detect more mutations in binding sites than RefSeq, or equivalently Uniprot and CDD, database annotations. However, this table also shows that including RefSeq annotations, as was done in our analysis, provides additional site information.

| Ligand | (Number of proteins with binding site inferred by homology)/(Number of proteins with RefSeq binding site) | Number of proteins with RefSeq but not homology-inferred binding sites | Number of proteins with homology-inferred but not RefSeq binding sites |
|---|---|---|---|
| $Ca^{2+}$ | 2348/1108 (2.1) | 246 | 1576 |
| $Cu^{2+}$ | 92/8 (12) | 0 | 84 |
| $Fe^{2+}$ | 79/33 (2.4) | 22 | 68 |
| $Mg^{2+}$ | 2643/309 (8.6) | 131 | 2465 |
| $Mn^{2+}$ | 618/4 (155) | 0 | 614 |
| $Zn^{2+}$ | 3131/1086 (2.9) | 401 | 2446 |
| Small molecules | 10123/5189 (2.0) | 1429 | 6363 |
| Proteins | 13585/5677 (2.4) | 1366 | 9274 |
| Nucleic acids | 3109/2108 (1.5) | 793 | 1794 |

**Table 4: Distribution of mutation clusters in oncogene and tumor suppressor proteins.** Only clusters that were significant with a false discovery rate (FDR) cutoff of 5% were included. Also only proteins with a minimum of ten mutations in the corresponding data set were considered.

| Data set | Oncogenes with clusters | Tumor suppressors with clusters | Other proteins with clusters | Fraction of proteins with clusters |
|---|---|---|---|---|
| COSMIC cancer mutations | 30 (27.2%) | 18 (16.4%) | 62 (56.4%) | 110/325 (33.8%) |
| HGMD | 10 (10.4%) | 8 (8.33%) | 78 (81.3%) | 96/562 (17.1%) |
| dbSNP neutral polymorphisms | 0 (0.0%) | 5 (0.0519%) | 19 (0.197%) | 24/9637 (0.249%) |
| Total | 496 | 874 | | |

**Table 5: Numbers of mutations in HGMD and dbSNP that are predicted to either disrupt or promote metal ion binding according to the multinomial model described in the text.**

| Metal ion | HGMD | | | dbSNP | | | |
|---|---|---|---|---|---|---|---|
| | $N_{dis}^{HGMD}$ = Number of mutations with p(mutant) < p(WT) | $N_{imp}^{HGMD}$ = Number of mutations with p(mutant) > p(WT) | * $N_{dis}^{HGMD} > N_{imp}^{HGMD}$ | $N_{dis}^{dbSNP}$ = Number of mutations with p(mutant) < p(WT) | $N_{imp}^{dbSNP}$ = Number of mutations with p(mutant) > p(WT) | * $N_{dis}^{dbSNP} > N_{imp}^{dbSNP}$ | † $\frac{N_{dis}^{HGMD}}{N_{imp}^{HGMD}} > \frac{N_{dis}^{dbSNP}}{N_{imp}^{dbSNP}}$ |
| $Ca^{2+}$ | 424 | 167 | Yes | 1795 | 982 | Yes | Yes |
| $Cu^{2+}$ | 28 | 3 | Yes | 28 | 28 | No | Yes |
| $Fe^{2+}$ | 16 | 3 | Yes | 24 | 12 | No | No |
| $Mg^{2+}$ | 249 | 134 | Yes | 890 | 523 | Yes | No |
| $Mn^{2+}$ | 42 | 8 | Yes | 204 | 66 | Yes | No |
| $Zn^{2+}$ | 320 | 108 | Yes | 2160 | 629 | Yes | No |

Significant at 5% level after Bonferroni correction using binomial (*) or Fisher exact test (†).

## Supplementary Information

**Figure S1: Divalent ion coordinating residue frequencies calculated from a non-redundant set of protein structures.**

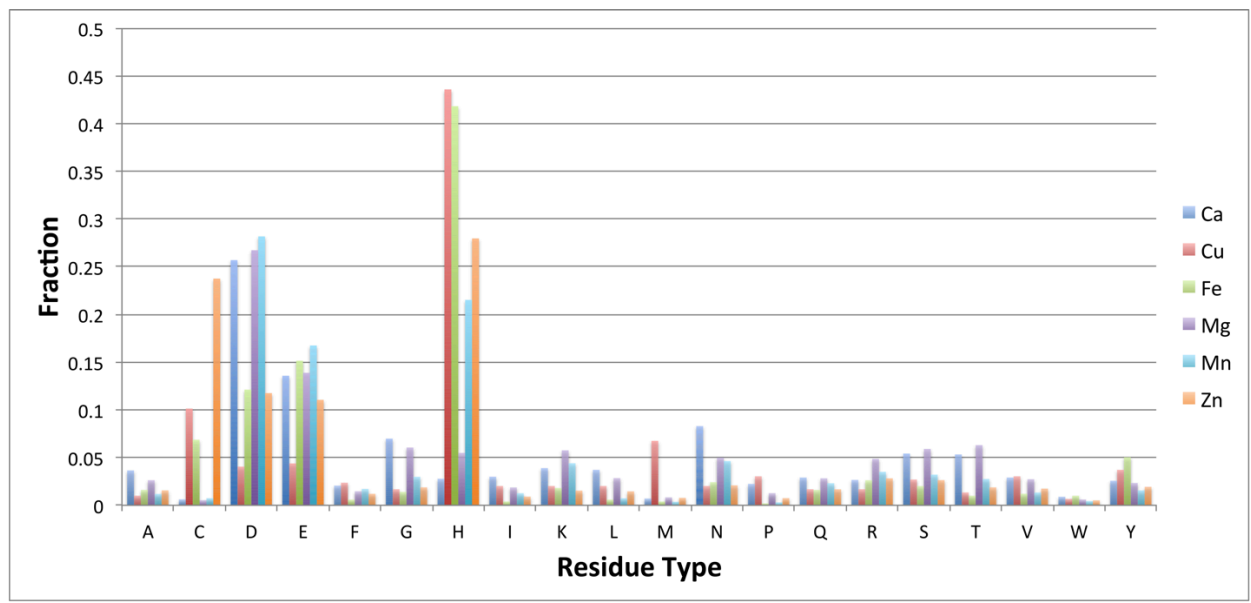

| Mutation type | disease | Functional | Cancer (COSMIC database) | | | | | Neutral | | |
|---|---|---|---|---|---|---|---|---|---|---|
| Mutation database | HGMD | Uniprot | All genes | MSKCC oncogenes | MSKCC tumor suppressors | Cancer Gene Census genes | CGP Cancer Cell Line Project | non-pathogenetic dbSNP | non-pathogenetic dbSNP (MAF ≥ 5%) | Uniprot humsavar.txt |
| Divalent metal ions | 1474 (4.10%) | 41 (3.22%) | 907 (3.15%) | 208 (5.05%) | 214 (3.89%) | 90 (4.52%) | 38 (14.1%) | 6810 (2.08%) | 122 (2.00%) | 217 (1.50%) |
| Small molecules | 5755 (16.0%) | 273 (21.5%) | 3553 (12.3%) | 967 (23.5%) | 976 (17.7%) | 373 (18.7%) | 69 (25.7%) | 30648 (9.34%) | 539 (8.86%) | 1583 (10.9%) |
| Proteins | 6082 (16.9%) | 243 (19.1%) | 4946 (17.2%) | 1239 (30.1%) | 1341 (24.3%) | 598 (30.1%) | 110 (40.9%) | 44636 (13.6%) | 975 (16.0%) | 2244 (15.4%) |
| Nucleic acids | 631 (1.75%) | 45 (3.54%) | 614 (2.13%) | 92 (2.23%) | 122 (2.21%) | 103 (5.18%) | 20 (7.43%) | 5238 (1.60%) | 78 (1.28%) | 153 (1.06%) |
| Any binding site* | 11278 (31.4%) | 484 (38.3%) | 8102 (28.1%) | 1794 (43.5%) | 2025 (36.8%) | 882 (44.3%) | 146 (54.3%) | 74133 (22.6%) | 1409 (23.1%) | 3459 (23.9%) |
| Potential small molecule binding sites (pockets) | 5755 (16.0%) | 498 (39.2%) | 9168 (31.8%) | 2158 (52.4%) | 2318 (42.1%) | 989 (49.7%) | 153 (56.9%) | 86037 (26.2%) | 1490 (24.5%) | 4137 (28.5%) |
| Predicted divalent metal ions | 1542 (4.29%) | 44 (3.46%) | 1011 (3.51%) | 287 (6.96%) | 268 (4.87%) | 163 (8.19%) | 58 (21.6%) | 8498 (2.59%) | 123 (2.02%) | 250 (1.72%) |
| Predicted proteins | 996 (2.77%) | 40 (3.14%) | 981 (3.40%) | 181 (4.39%) | 326 (5.92%) | 69 (3.47%) | 33 (12.3%) | 6957 (2.12%) | 125 (2.05%) | 277 (1.91%) |
| Total number of mutations in the data set† | 35966 | 1272 | 28816 | 4122 | 5508 | 1990 | 269 | 328152 | 6085 | 14494 |
| Destabilizing‡ | 4766 (22.4%) | | 2006 (10.7%) | 404 (12.7%) | 528 (13.4%) | 194 (11.9%) | | 17688 (8.48%) | 366 (9.95%) | 611 (4.20%) |
| Total number of mutations with stability values | 18724 | | 18809 | 3171 | 3935 | 1627 | | 208491 | 3677 | 14548 |

**Table S1: Numbers of variants in each data set occurring in specific classes of binding sites.** The percentages for binding sites inferred by homology or predicted by machine learning are plotted in Figures 3 and 4, respectively.

*Includes metal ion, small molecule, protein, and nucleic acid binding sites inferred by homology.

†Only mutations mapped to at least one protein structure are considered.

‡Destabilizing mutations are defined to have predicted ΔΔG > 2.0 kcal/mol. Percentages are calculated relative to the number of mutations with available ΔΔG values shown in the bottom row.

**Table S2: Input features used for the Random Forest classification of functional and neutral mutations.**

| Description | Type | Number of variables |
|---|---|---|
| In divalent metal ion, protein, small molecule, or nucleic acid binding site inferred by homology | binary | 9 |
| In predicted divalent metal ion or protein binding site | binary | 2 |
| In surface pocket | binary | 1 |
| RefSeq ion, polypeptide, nucleotide, chemical, or other binding site | binary | 5 |
| In dbPTM [47] site | binary | 1 |
| Median relative SASA | real | 1 |
| Median absolute SASA | real | 1 |
| Consensus DSSP secondary structure classification | real | 1 |
| Median B-factor | real | 1 |
| Predicted change in protein stability (ΔΔG) | real | 1 |
| Wild-type and mutant residue type | binary | 40 |

**Table S3: Divalent ion coordinating residue frequencies.**

|    | A | C | D | E | F | G | H | I | K | L | M | N | P | Q | R | S | T | V | W | Y |
|----|---|---|---|---|---|---|---|---|---|---|---|---|---|---|---|---|---|---|---|---|
| Ca | 3.657E-02 | 6.295E-03 | 2.569E-01 | 1.358E-01 | 2.068E-02 | 6.984E-02 | 2.788E-02 | 2.998E-02 | 3.897E-02 | 3.717E-02 | 7.194E-03 | 8.303E-02 | 2.248E-02 | 2.908E-02 | 2.668E-02 | 5.426E-02 | 5.336E-02 | 2.908E-02 | 8.993E-03 | 2.578E-02 |
| Cu | 1.014E-02 | 1.014E-01 | 4.054E-02 | 4.392E-02 | 2.365E-02 | 1.689E-02 | 4.358E-01 | 2.027E-02 | 2.027E-02 | 2.027E-02 | 6.757E-02 | 2.027E-02 | 3.041E-02 | 1.689E-02 | 1.689E-02 | 2.703E-02 | 1.351E-02 | 3.041E-02 | 6.757E-03 | 3.716E-02 |
| Fe | 1.616E-02 | 6.869E-02 | 1.212E-01 | 1.515E-01 | 6.061E-03 | 1.414E-02 | 4.182E-01 | 4.040E-03 | 1.818E-02 | 6.061E-03 | 4.040E-03 | 2.424E-02 | 2.020E-03 | 1.616E-02 | 2.626E-02 | 2.020E-02 | 1.010E-02 | 1.212E-02 | 1.010E-02 | 5.051E-02 |
| Mg | 2.634E-02 | 5.415E-03 | 2.673E-01 | 1.388E-01 | 1.477E-02 | 6.055E-02 | 5.489E-02 | 1.871E-02 | 5.759E-02 | 2.855E-02 | 8.368E-03 | 4.996E-02 | 1.280E-02 | 2.830E-02 | 4.873E-02 | 5.907E-02 | 6.301E-02 | 2.732E-02 | 6.153E-03 | 2.338E-02 |
| Mn | 1.196E-02 | 7.474E-03 | 2.818E-01 | 1.674E-01 | 1.719E-02 | 2.990E-02 | 2.152E-01 | 1.271E-02 | 4.410E-02 | 7.474E-03 | 3.737E-03 | 4.634E-02 | 2.990E-03 | 2.317E-02 | 3.513E-02 | 3.214E-02 | 2.765E-02 | 1.345E-02 | 4.484E-03 | 1.570E-02 |
| Zn | 1.571E-02 | 2.375E-01 | 1.176E-01 | 1.105E-01 | 1.205E-02 | 1.885E-02 | 2.797E-01 | 9.165E-03 | 1.545E-02 | 1.466E-02 | 7.855E-03 | 2.095E-02 | 7.594E-03 | 1.676E-02 | 2.828E-02 | 2.645E-02 | 1.885E-02 | 1.754E-02 | 5.237E-03 | 1.938E-02 |

**Table S4: PDB heterocompounds that were excluded from the set of small molecules.**
These compounds are either crystallization additives or ions not included in the separate divalent metal ion set.

| PDB Heterocompound ID | Compound Name |
|---|---|
| F | FLUORIDE ION |
| K | POTASSIUM ION |
| AG | SILVER ION |
| AL | ALUMINUM ION |
| BA | BARIUM ION |
| BR | BROMIDE ION |
| CD | CADMIUM ION |
| CL | CHLORIDE ION |
| CM | CARBOXYMETHYL GROUP |
| CN | CYANIDE |
| CO | COBALT (II) ION |
| CS | CESIUM ION |
| HG | MERCURY (II) ION |
| LI | LITHIUM ION |
| NA | SODIUM ION |
| NI | NICKEL (II) ION |
| OH | HYDROXIDE ION |
| PB | LEAD (II) ION |
| RB | RUBIDIUM ION |
| SR | STRONTIUM ION |
| Y1 | YTTRIUM ION |
| 12P | DODECAETHYLENE GLYCOL |
| 144 | TRIS-HYDROXYMETHYL-METHYL-AMMONIUM |
| 15P | POLYETHYLENE GLYCOL (N=34) |
| 16D | HEXANE-1,6-DIAMINE |
| 1BO | 1-BUTANOL |
| 1PS | 3-PYRIDINIUM-1-YLPROPANE-1-SULFONATE |
| 2OS | 3-N-OCTANOYLSUCROSE |
| 2PE | NONAETHYLENE GLYCOL |
| 3CO | COBALT (III) ION |
| 3NI | NICKEL (III) ION |
| ACA | 6-AMINOHEXANOIC ACID |
| ACN | ACETONE |
| ACT | ACETATE ION |
| ACY | ACETIC ACID |
| AGC | ALPHA-D-GLUCOSE |
| AZI | AZIDE ION |
| B3P | 2-[3-(2-HYDROXY-1,1-DIHYDROXYMETHYL-ETHYLAMINO)-PROPYLAMINO]-2-HYDROXYMETHYL-PROPANE-1,3-DIOL |
| B7G | HEPTYL-BETA-D-GLUCOPYRANOSIDE |
| BCN | BICINE |
| BCT | BICARBONATE ION |
| BE7 | (4-CARBOXYPHENYL)(CHLORO)MERCURY |
| BEQ | N-(CARBOXYMETHYL)-N,N-DIMETHYL-3-[(1-OXODODECYL)AMINO]-1-PROPANAMINIUM INNER SALT |
| BGC | BETA-D-GLUCOSE |
| BMA | BETA-D-MANNOSE |
| BNG | B-NONYLGLUCOSIDE |
| BOG | B-OCTYLGLUCOSIDE |
| BRO | BROMO GROUP |
| BTB | 2-[BIS-(2-HYDROXY-ETHYL)-AMINO]-2-HYDROXYMETHYL-PROPANE-1,3-DIOL |
| BTC | CYSTEINE |
| BU1 | 1,4-BUTANEDIOL |
| BU2 | 1,3-BUTANEDIOL |
| BU3 | (R,R)-2,3-BUTANEDIOL |
| C10 | HEXAETHYLENE GLYCOL MONODECYL ETHER |
| C8E | (HYDROXYETHYLOXY)TRI(ETHYLOXY)OCTANE |
| CAC | CACODYLATE ION |
| CBM | CARBOXYMETHYL GROUP |
| CBX | CARBOXY GROUP |
| CCN | ACETONITRILE |
| CE1 | O-DODECANYL OCTAETHYLENE GLYCOL |
| CIT | CITRIC ACID |
| CLO | CHLORO GROUP |
| CM5 | 5-CYCLOHEXYL-1-PENTYL-BETA-D-MALTOSIDE |
| CPS | 3-[(3-CHOLAMIDOPROPYL)DIMETHYLAMMONIO]-1-PROPANESULFONATE |
| CRY | PROPANE-1,2,3-TRIOL |
| CXE | PENTAETHYLENE GLYCOL MONODECYL ETHER |
| CYN | CYANIDE ION |
| CYS | CYSTEINE |
| DDQ | DECYLAMINE-N,N-DIMETHYL-N-OXIDE |
| DHD | 2,4-DIOXO-PENTANEDIOIC ACID |
| DIA | OCTANE 1,8-DIAMINE |
| DIO | 1,4-DIETHYLENE DIOXIDE |
| DMF | DIMETHYLFORMAMIDE |
| DMS | DIMETHYL SULFOXIDE |
| DMU | DECYL-BETA-D-MALTOPYRANOSIDE |
| DMX | 3-[BENZYL(DIMETHYL)AMMONIO]PROPANE-1-SULFONATE |
| DOD | DEUTERATED WATER |
| DOX | DIOXANE |
| DPR | D-PROLINE |
| DR6 | ALPHA-[4-(1,1,3,3 - TETRAMETHYLBUTYL)PHENYL]-OMEGA-HYDROXY-POLY(OXY-1,2-ETHANEDIYL) |
| DXG | 4-DEOXYGLUCARATE |
| EDO | 1,2-ETHANEDIOL |
| EEE | ETHYL ACETATE |
| EGL | ETHYLENE GLYCOL |
| EOH | ETHANOL |
| ETF | TRIFLUOROETHANOL |
| FCL | 3-CHLORO-L-PHENYLALANINE |
| FCY | FREE CYSTEINE |
| FLO | FLUORO GROUP |
| FMT | FORMIC ACID |
| FRU | FRUCTOSE |
| GBL | GAMMA-BUTYROLACTONE |
| GCD | 4,5-DEHYDRO-D-GLUCURONIC ACID |
| GLC | ALPHA-D-GLUCOSE |
| GLO | D-GLUCOSE IN LINEAR FORM |
| GLY | GLYCINE |
| GOL | GLYCEROL |
| GPX | GUANOSINE 5'-DIPHOSPHATE 2':3'-CYCLIC MONOPHOSPHATE |
| HEZ | HEXANE-1,6-DIOL |
| HTG | HEPTYL 1-THIOHEXOPYRANOSIDE |
| HTO | HEPTANE-1,2,3-TRIOL |
| ICI | ISOCITRIC ACID |
| ICT | ISOCITRIC ACID |
| IDO | IODO GROUP |
| IDT | 4,5-DEHYDRO-L-IDURONIC ACID |
| IOD | IODIDE ION |
| IOH | 2-PROPANOL, ISOPROPANOL |
| IPA | ISOPROPYL ALCOHOL |
| IPH | PHENOL |
| JEF | O-(O-(2-AMINOPROPYL)-O'-(2-METHOXYETHYL)POLYPROPYLENE GLYCOL 500) |
| LAK | BETA-D-GALACTOPYRANOSYL-1-6-BETA-D-GLUCOPYRANOSE |
| LAT | BETA-LACTOSE |
| LBT | ALPHA-LACTOSE |
| LDA | LAURYL DIMETHYLAMINE-N-OXIDE |
| LMT | DODECYL-BETA-D-MALTOSIDE |
| MA4 | CYCLOHEXYL-HEXYL-BETA-D-MALTOSIDE |
| MAN | ALPHA-D-MANNOSE |
| MG8 | N-OCTANOYL-N-METHYLGLUCAMINE |
| MHA | (CARBAMOYLMETHYL-CARBOXYMETHYL-AMINO)-ACETIC ACID |
| MOH | METHANOL |
| MPD | (4S)-2-METHYL-2,4-PENTANEDIOL |
| MPO | 3[N-MORPHOLINO]PROPANE SULFONIC ACID |
| MRD | (4R)-2-METHYLPENTANE-2,4-DIOL |
| MRY | MESO-ERYTHRITOL |
| MSE | SELENOMETHIONINE |
| MTL | D-MANNITOL |
| N8E | 3,6,9,12,15-PENTAOXATRICOSAN-1-OL |
| NCO | COBALT HEXAMMINE(III) |
| NH4 | AMMONIUM ION |
| NHE | 2-[N-CYCLOHEXYLAMINO]ETHANE SULFONIC ACID |
| NO3 | NITRATE ION |
| OTE | 2-{2-[2-(2-OCTYLOXY-ETHOXY)-ETHOXYL]-ETHOXY}ETHANOL |
| P33 | 3,6,9,12,15,18-HEXAOXAICOSANE-1,20-DIOL |
| P4C | O-ACETALDEHYDYL-HEXAETHYLENE GLYCOL |
| PDO | 1,3-PROPANDIOL |
| PE4 | 2-{2-[2-(2-{2-[2-(2-ETHOXY-ETHOXY)-ETHOXY]-ETHOXY}-ETHOXY)-ETHOXY]-ETHOXY}-ETHANOL |
| PE7 | 1-DEOXY-1-THIO-HEPTAETHYLENE GLYCOL |
| PE8 | 3,6,9,12,15,18,21-HEPTAOXATRICOSANE-1,23-DIOL |
| PEU | 2,5,8,11,14,17,20,23,26,29,32,35,38,41,44,47,50,53,56,59,62,65,68,71,74,77,80-HEPTACOSAOXADOOCTACONTAN-82-OL |
| PG5 | 1-METHOXY-2-[2-(2-METHOXY-ETHOXY]-ETHANE |
| PG6 | 1-(2-METHOXY-ETHOXY)-2-{2-[2-(2-METHOXY-ETHOXY]-ETHOXY}-ETHANE |
| PGE | TRIETHYLENE GLYCOL |
| PGO | S-1,2-PROPANEDIOL |
| PGQ | S-1,2-PROPANEDIOL |
| PGR | R-1,2-PROPANEDIOL |
| PIG | 2-[2-(2-HYDROXY-ETHOXY)-ETHOXY]-ETHANOL |
| PIN | PIPERAZINE-N,N'-BIS(2-ETHANESULFONIC ACID) |
| PO4 | PHOSPHATE ION |
| POL | N-PROPANOL |
| SAL | 2-HYDROXYBENZOIC ACID |
| SBT | 2-BUTANOL |
| SCN | THIOCYANATE ION |
| SDS | DODECYL SULFATE |
| SO4 | SULFATE ION |
| SOR | D-SORBITOL |
| SPD | SPERMIDINE |
| SPK | SPERMINE (FULLY PROTONATED FORM) |
| SPM | SPERMINE |
| SUC | SUCROSE |
| SUL | SULFATE ANION |
| TAR | D(-)-TARTARIC ACID |
| TAU | 2-AMINOETHANESULFONIC ACID |
| TBU | TERTIARY-BUTYL ALCOHOL |
| TEP | THEOPHYLLINE |
| TFP | 10-[3-(4-METHYL-PIPERAZIN-1-YL)-PROPYL]-2-TRIFLUOROMETHYL-10H-PHENOTHIAZINE |
| TLA | L(+)-TARTARIC ACID |
| TMA | TETRAMETHYLAMMONIUM ION |
| TRE | TREHALOSE |
| TRS | 2-AMINO-2-HYDROXYMETHYL-PROPANE-1,3-DIOL |
| TRT | FRAGMENT OF TRITON X-100 |
| UMQ | UNDECYL-MALTOSIDE |
| UNX | UNKNOWN ATOM OR ION |
| URE | UREA |
| XPE | 3,6,9,12,15,18,21,24,27-NONAOXANONACOSANE-1,29-DIOL |
| YT3 | YTTRIUM (III) ION |